\documentclass[twocolumn,showpacs,prb]{revtex4}

\usepackage{graphicx}
\usepackage{dcolumn}
\usepackage{bm}
\usepackage{amssymb}

\begin{document}


\title{Low dimensional ordering and fluctuations in methanol-$\beta$-hydroquinone-clathrate studied by X-ray and neutron diffraction}

\author{Maikel~C.~Rheinst\"adter}
\altaffiliation[Present address: ]{Institut Laue-Langevin, 6 rue
Jules Horowitz, BP 156, 38042 Grenoble Cedex 9, France}
\email{rheinstaedter@ill.fr} \affiliation{Institut Laue-Langevin,
6 rue Jules Horowitz, BP 156,38042 Grenoble Cedex 9, France}
\affiliation{Technische Physik, Universit\"at des Saarlandes, PSF
1551150, 66041 Saarbr\"ucken, Germany}

\author{Mechthild~Enderle}
\affiliation{Institut Laue-Langevin, 6 rue Jules Horowitz, BP
156,38042 Grenoble Cedex 9, France}

\author{Axel~Kl\"opperpieper}
\affiliation{Technische Physik, Universit\"at des Saarlandes, PSF
1551150, 66041 Saarbr\"ucken, Germany}

\author{Klaus~Knorr}
\affiliation{Technische Physik, Universit\"at des Saarlandes, PSF
1551150, 66041 Saarbr\"ucken, Germany}

\date{\today}

\begin{abstract}
Methanol-$\beta$-hydroquinone-clathrate has been established as a
model system for dielectric ordering and fluctuations and is
conceptually close to magnetic spin systems. In X-ray and neutron
diffraction experiments, we investigated the ordered structure,
the one-dimensional (1D) and the three-dimensional (3D) critical
scattering in the paraelectric phase, and the temperature
dependence of the lattice constants. Our results can be explained
by microscopic models of the methanol pseudospin in the
hydroquinone cage network, in consistency with previous dielectric
investigations.
\end{abstract}



\maketitle

Reducing the dimensionality has a drastic influence on
order-disorder phenomena, as it generally enhances fluctuations.
Phase transitions are shifted to lower temperatures, or may be
suppressed entirely. In the one-dimensional Ising model with
nearest-neighbor interaction, thermal fluctuations suppress
ordering at any finite temperature, independent of the strength of
the interaction. Theoretical models for low-dimensional ordering
and fluctuations are mostly tested on magnetically interacting
materials, where the magnetic moment is well decoupled from the
surrounding chemical lattice, and the interaction is short-ranged.
Materials with low-dimensional, but long-ranged interactions like
the dipole-dipole interaction are rare. Due to its small energy
scale, the magnetic dipole-dipole interaction is often negligible
compared to exchange-type interactions, magneto-elastic
interactions and anisotropies induced by spin-orbit coupling and
crystal field. The electric dipole-dipole interaction (EDD) is
several orders of magnitude larger than its magnetic pendant. The
electric dipole moment is connected to polar molecules or to
displacements of groups of atoms against each other. Usually, the
interaction with the surrounding lattice is far from being
negligible. Materials, which allow to test theoretical predictions
for fluctuations and ordering of ''pseudospins'' interacting
exclusively via the EDD interaction, are extremely rare. In this
context, clathrates with polar guest molecules provide a happy
exception. They may undergo structural phase transitions driven by
the interactions between the dipole moments of the guest
molecules, with practically no coupling of the orientational
degrees of freedom to the host lattice.

The present article deals with methanol-$\beta$-hydroquinone
clathrate, where well defined methanol electric dipoles occupy the
almost spherical hydroquinone cavities. The electric dipoles
arrange in chains and are well decoupled from the surrounding host
lattice. Thus the methanol clathrate can be understood in terms of
interacting point-dipoles and has been established as a model
system for dipolar ordering processes
\cite{Woll:2000,Woll:2001,Kityk:2002,Rheinstaedter:2003,Rheinstaedter:2004}.

\begin{figure}
\centering
\resizebox{1.00\columnwidth}{!}{\rotatebox{0}{\includegraphics{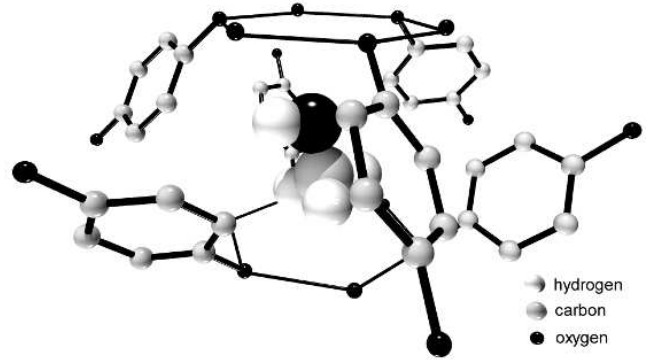}}}
\caption[]{The methanol molecule (CH$_3$OH) in the quinone cage
network. The atoms of the guest molecules are shown with their
Van-der-Waals diameters whereas for the hudroquinone host lattice
the stich-and-ball presentation is used.}
\label{clathratkaefig_mit_methanol_020111.eps}
\end{figure}

Figure \ref{clathratkaefig_mit_methanol_020111.eps} shows the
methanol molecule, which carries a dipole moment of 1.69~Debye
\cite{Burkhard:1951}, in the hydroquinone cage. The shortest
distance between the guest molecules is along the hexagonal
c-axis. The geometry of the lattice combined with EDD interaction
is responsible for the formation of ferroelectric chains along the
c-axis, reflected in the dielectric properties: Woll {\em et
al.\@} have studied the dielectric response as a function of
temperature and frequency \cite{Woll:2000,Woll:2001}. The static
dielectric constant along the hexagonal $c$-axis,
$\varepsilon_{cs}$, shows a strong deviation from the Curie-Weiss
behavior already at 250~K. The temperature dependence of
$\varepsilon_{cs}$ at higher temperatures is well described by a
quasi one-dimensional (1D) Ising model with a ferroelectric
coupling $J_c$ to nearest neighbors within the chain and a
mean-field antiferroelectric inter-chain interaction with coupling
constant $J_{\perp}$ \cite{Scalapino:1975}. At lower T,
$\varepsilon_{cs}(T)$ deviates from this mean-field-Ising-behavior
and finally decreases with decreasing T, indicating the onset of
three-dimensional (3D) antiferroelectric inter-chain correlations.
At T$_c$=65~K the methanol-clathrate undergoes a transition into
an antiferroelectrically ordered structure.

While the dielectric experiments probe the macroscopic response at
$Q=0$, at the center of the Brillouin zone, diffraction
experiments give access to the corresponding underlying
microscopic processes and allow to study the structural
susceptibility at arbitrary points in reciprocal space. In this
article, we first present X-ray diffraction results obtained in
the ordered phase, which allow to develop the dipolar structure
model. Then we turn to the paraelectric regime, where we studied
the 1D and 3D critical scattering by X-ray and neutron diffraction
experiments, as well as the temperature dependence of the lattice
parameters. We compare our results to theoretical predictions and
discuss them in the context of the preceding dielectric
experiments.


\section*{Experimental}\label{Experimental}
Single crystals of methanol-$\beta$-hydroquinone clathrate have
been grown from a saturated solution of quinol and methanol at 313
K. Completely deuterated samples were used for neutron experiments
to minimize the incoherent background. Typical sample crystals
were 2x2x2 mm$^3$. At room temperature, the methanol-clathrate has
a rhombohedral R$\bar{3}$ structure \cite{PalinPowell:1948b}.
Throughout this article, we use the hexagonal notation, which is
better adapted to the dielectric properties of the clathrate than
the rhombohedral. Exceptions are marked explicitly by
$(\ldots)_{rh}$. The hexagonal lattice parameters at 300~K are
$a$=16.62~\AA, $c$=5.56~\AA. For the X-ray investigations, we used
monochromatic CuK$_{\alpha}$ radiation and a two-circle
diffractometer, equipped with a closed cycle refrigerator. The
single crystalline samples were oriented with $c$ perpendicular to
the scattering plane. This allows scans within the $(hk0)$ plane.
Scans within planes $(hkl)$, $l$ constant, $l\leq 1$ are possible
by moving the detector out of the diffractometer plane. Rotating
crystal exposures have also been taken. Here the crystal was
rotated about $c$ and the incoming beam was perpendicular to the
$c$-axis. Diffuse 1D critical scattering has been measured at the
two-circle neutron diffractometer D23 (CEA-CRG) at the high flux
reactor of the ILL in Grenoble, France. Here the 1D axis, $c$, was
mounted in the scattering plane.

\section*{Results}\label{Results}
\subsection*{Ordered phase}
\begin{figure}
\centering
\resizebox{1.00\columnwidth}{!}{\rotatebox{0}{\includegraphics{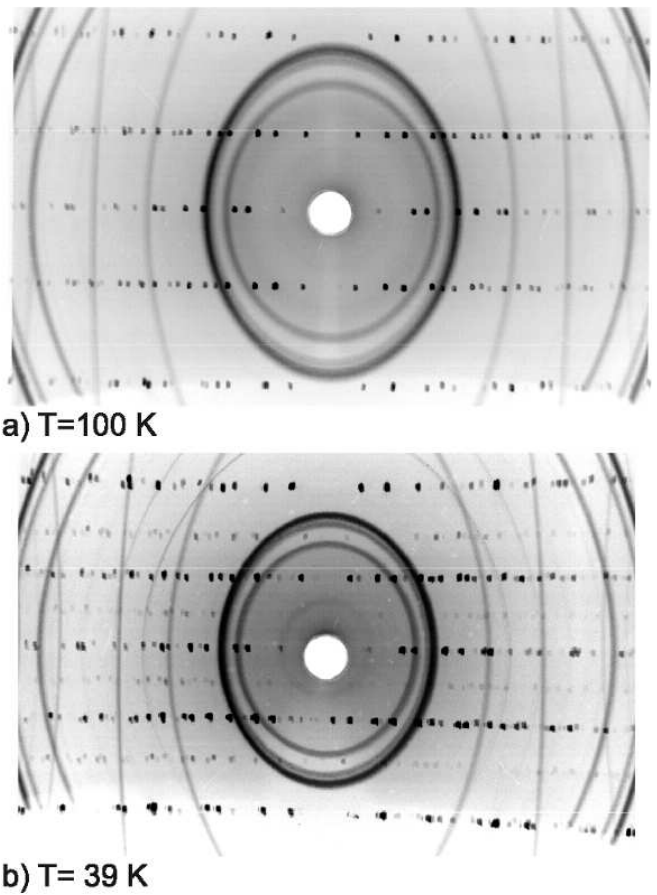}}}
\caption[]{Rotating crystal photographs taken at T=100~K (a) and
39~K (b). The elliptical lines are due to scattering from the
Al-windows of the cryo-refrigerator.} \label{rotatingcrystal}
\end{figure}

Rotating-crystal X-ray photographs of methanol-clathrate were
taken above and below the phase transition ($T_c=65$~K). Figure
\ref{rotatingcrystal}a) displays the photograph taken in the
paraelectric phase at T=100~K, figure \ref{rotatingcrystal}b) in
the ordered phase, at 39~K. At high temperatures, horizontal
layers of reflections $(hkl)$ with $l=0, \pm 1,\pm 2,\ldots$ are
visible.
Below T$_c$, additional reflections in layers with integer and
half-integer values of $l$, $l=\pm\frac{1}{2},
\pm\frac{3}{2},\ldots$ appear. The $(hkl)$ values of the
superlattice reflections can be determined from their position on
the photograph. Careful analysis shows that modes with wave
vectors $k_1=(00\frac{1}{2})_{rh}$ and
$k_2=(\frac{1}{2}\frac{1}{2}\frac{1}{2})_{rh}$ condense at the
phase transition on the boundary of the rhombohedral Brillouin
zone at the Lifshitz points F ($k_1$) and L ($k_2$). To generate
all superlattice reflections as linear combinations of $k_1$ and
$k_2$, L and two arms of the F-star are required.
This means, two of the rhombohedral lattice vectors double at
T$_c$. Additionally, the unit cell doubles in the hexagonal $c$
direction (as indicated by the L-type vector
$k_2=(\frac{1}{2}\frac{1}{2}\frac{1}{2})_{rh})$.

\begin{figure} \centering
\resizebox{1.00\columnwidth}{!}{\rotatebox{0}{\includegraphics{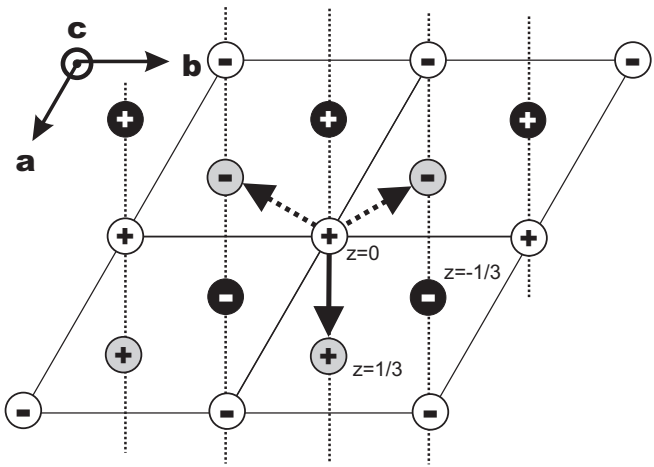}}}
\caption[]{Projection of the 3D ordered antiferroelectric
structure onto the hexagonal basal plane: white circles: dipoles
at $z=0$, grey circles: $z=\frac{1}{3}$, black circles:
$z=-\frac{1}{3}$. +/- indicates the orientation of the electric
dipole. Below T$_c$, two of the rhombohedral lattice vectors
double (indicated by the dotted arrows). Along $c$, the dipoles
order ferroelectrically. The ferroelectric chains arrange in
ferro-ordered (100) layers with alternating sign from layer to
layer. These layers are visualized by dotted lines. Solid lines
indicate the hexagonal unit cells.}
\label{methanol_kettenstruktur_sw.eps}
\end{figure}

The occurrence of $k_1$ indicates the freezing of the methanol
molecules into one of the six symmetry directions of the
R$\bar{3}$ space group. The resulting structure of the dipole
system in the dielectrically ordered phase is shown in
Fig.~\ref{methanol_kettenstruktur_sw.eps}.
The dipoles are arranged in ferroelectrically ordered chains
running along the hexagonal $c$-axis, which are arranged in sheets
of alternating sign. As a property of the EDD interaction and the
lattice symmetry, the dipoles prefer an antiferroelectric
arrangement perpendicular to the chains, which can not be
satisfied for all chains, because of the triangular lattice
symmetry. As a consequence, every dipole chain is surrounded by 4
chains, which are oriented antiparallel, and by 2 parallel
oriented dipole chains. The antiferroelectric interaction is thus
satisfied by the majority. The 4:2 pattern fills the hexagonal
basal plane in a regular fashion: Every dipole has the same
environment of ferro- and antiferroelectrically oriented
pseudospins. This 2-$k_1$ ground state is also obtained from Monte
Carlo calculations \cite{Sixou:1976} and simple lattice sum
calculations, considering the EDD interaction of the guest
molecules, only (see Ref.~[\onlinecite{Rheinstaedter:2002}] for
the case of acetonitrile).

The doubling of the unit cell along $c$, evidenced by the
occurrence of L-type $k_2$ reflections, indicates an antiphase
arrangement in neighboring cells along $c$. $k_2$ indicates a
dimerization of the dipoles in the 1D chain, i.e.\@ the appearance
of pairs of strongly coupled pseudospins. Considering the EDD
interaction in combination with a flat potential in the cage
center, the latter dimerization is energetically favorable, as we
will discuss below. The ordering of the dipole system at T$_c$ is
connected with a structural phase transition into the triclinic
space group P1, connected with a small triclinic distortion of the
host lattice below T$_c$ \cite{Woll:2001}.

\subsection*{Paraelectric regime}

\begin{figure*} \centering
\resizebox{0.75\textwidth}{!}{\rotatebox{0}{\includegraphics{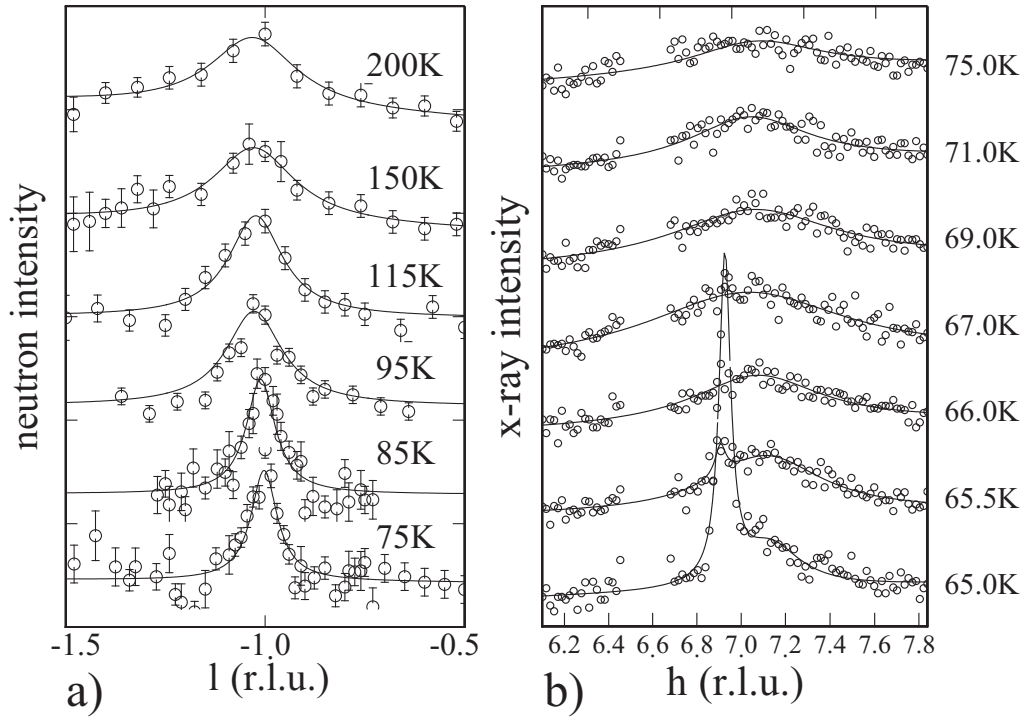}}}
\caption[]{(a): Diffuse 1D-scattering at
Q=$(3\;\overline{4.5}\;\overline{1}+\xi)$ (neutron measurement)
and (b) diffuse 3D-scattering at
Q=$(7+\xi\;\overline{3}-0.43\xi\;\overline{0.5})$ (measured with
X-rays). At 65~K the superlattice reflection of the ordered low-T
phase is already visible as a sharp peak. (An Al-powder line has
been cut around $h=6.55$.)} \label{1dpics.eps}
\end{figure*}

\begin{figure} \centering
\resizebox{1.00\columnwidth}{!}{\rotatebox{0}{\includegraphics{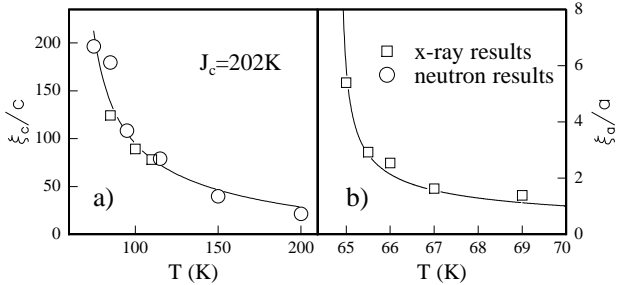}}}
\caption[]{Correlation length $\xi_c$ (a) und $\xi_{\perp}$ (b),
parallel and perpendicular to the dipole chains as it was
determined from Fig.~\ref{1dpics.eps}. Solid lines are fits after
Eq.~(\ref{korrelationslaenge1}) and (\ref{korrelationslaenge2}).
Results from neutron diffraction are marked by ($\circ$), X-ray
results by ($\square$).} \label{kohaerpics}
\end{figure}

The 1D fluctuations in the dipole chains give rise to diffuse
scattering which is distributed in planes perpendicular to the 1D
axis (and weighted by the methanol molecular form factor). The
onset of 3D ordering close to the phase transition is indicated in
diffuse scattering at the positions of the superlattice
reflections (of the low-T phase). The correlation length along the
dipole chains, $\xi_c$, and perpendicular, $\xi_{\perp}$, follows
from the quasi 1D Ising model, which has also been applied to fit
the dielectric data, to \cite{Scalapino:1975}:
\begin{eqnarray}
\left(\frac{\xi_c}{c}\right)&=&\left(\frac{T_c}{2J_c}\right)^{(1/2)}\frac{e^{2J_c/T_c}}{2|t|^{(1/2)}}\hspace{0.25cm}\label{korrelationslaenge1}\\
&\mbox{and}&\nonumber\\
\left(\frac{\xi_{\perp}}{a}\right)&=&\left(\frac{T_c}{4J_c}\right)^{(1/2)}\frac{1}{|t|^{(1/2)}}
\label{korrelationslaenge2}
\end{eqnarray}
($t=(T-T_c)/T_c$ is the reduced critical temperature). Figure
\ref{1dpics.eps} shows diffuse 1D and 3D scattering as measured
with X-ray and neutron diffraction at different temperatures in
the paraelectric regime. We investigated the weak 1D scattering on
a neutron diffractometer because the contribution of the methanol
molecules to the total scattering is higher in neutron scattering
with deuterated samples than in the X-ray experiments.
The correlation lengths $\xi_c$ and $\xi_{\perp}$ follow from the
width of Lorentzian peak profiles and are shown in Figure
\ref{kohaerpics}. The solid lines in the figure are fits after
Eq.~(\ref{korrelationslaenge1}) and (\ref{korrelationslaenge2}).
The fitting parameter $J_c$ is determined to $J_c=202$ K, in
excellent agreement with the dielectric results (195 K
\cite{Woll:2000}). The structural data are well described by the
quasi 1D Ising model.
The onset of the 3D correlations close to T$_c$ finally leads to
the phase transition into a 3D antiferroelectrically ordered
structure.

There is no indication of a static dimerization of the dipole
chains in the paraelectric phase, i.e.\@ no intensity at the
superlattice positions of $k_2$. But as a true 1D feature, the
dimerization could already appear in the dynamically ordered chain
segments in the paraelectric regime, where the 1D fluctuations
already reach some hundreds of unit cells.
We found no
sign of diffuse scattering in planes with half-integer values of
$l$, as caused by fluctuations of the dimerization above T$_c$.
From calculations of the methanol structure factor \footnote{To
calculate the form factor we assumed a tilt of the elementary
dipole of 22 degs to the hexagonal $c$-axis, as determined from
the dielectric experiments in
Refs.~[\onlinecite{Woll:2000,Woll:2001}].} where we allowed a
displacement of the dipoles along the 1D axis of
$\delta=0.25$~\AA, the intensity of the diffuse scattering due to
the dimerization can be estimated as being at least two orders of
magnitude smaller than that of the 1D critical scattering. The
corresponding correlation length $\xi_{sp}$ should be smaller than
$\xi_c$, the corresponding critical scattering as a consequence
broader. If the dimerization of the dipole chains already occurs
in the paraelectric phase, this effect presumably is too small to
be seen in standard diffraction experiments.

\begin{figure}
\centering
\resizebox{1.00\columnwidth}{!}{\rotatebox{0}{\includegraphics{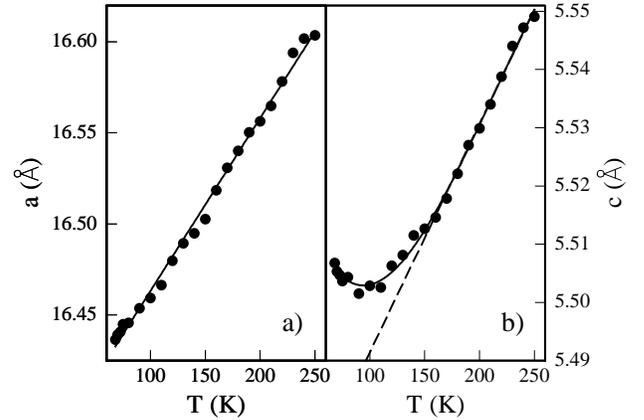}}}
\caption[]{Temperature dependence of the hexagonal lattice
parameters $a$ (a) und $c$ (b) of the hexagonal cell. The dotted
line in part (b) is a linear extrapolation of the high temperature
linear behavior to lower T.} \label{gitterkonstanten}
\end{figure}

Figure \ref{gitterkonstanten} shows the temperature dependence of
the hexagonal lattice parameters $a$ and $c$, as derived from the
Bragg positions. Whilst the $a$-axis decreases linearly with
temperature, $c$ shows a distinct anomaly below about 170~K.


\section*{Discussion}\label{Discussion}
The structure of the dipole system in the dielectrically ordered
phase in methanol-$\beta$-hydroquinone-clathrate, as determined
from the analysis of the superlattice reflections below T$_c$,
agrees well with the ground state calculated by Monte Carlo
simulations \cite{Sixou:1976} and lattice sum calculations. The
temperature behavior of the correlation length $\xi_c$ and
$\xi_{\perp}$ is well described by the corresponding quasi 1D
model by Scalapino {\em et al.\@} \cite{Scalapino:1975}. The
structural investigations therefore confirm the microscopic
picture for low dimensional ordering and fluctuations of the
methanol pseudospins in the $\beta$-hydroquinone cagework, as it
was deduced from the dielectric experiments.

An interesting point is the appearance of superlattice reflections
of type ''$k_2$'', because they indicate a doubling of the unit
cell in the 1D direction. The most probable explanation of this
doubling is a dimerization of the dipoles in the 1D chains.
Considering a 1D chain of electric dipoles on a deformable lattice
(with distance $d$), interacting solely by the electric
dipole-dipole interaction, a dimerization of the chain with
respect to an alternating distance $d+\delta$ and $d-\delta$ of
the dipoles is energetically favorable. Because of the rather
steep $r^{-3}$ potential, a dipole can lower its energy much more
by approaching its neighbor by a distance $\delta$ than it at the
same time looses potential energy by removing from the second
neighbor in the 1D chain.
The small lattice distortion at T$_c$ (a first order structural
transition) finally leads to a coupling of the two otherwise
independent wave vectors $k_1$ and $k_2$. The freezing of the
methanol-molecule and the static dimerization of the pseudospins
along the chains therefore fall together at T$_c$. In a system
with smaller coupling of the pseudospins to the surrounding host
lattice, two successive transitions should occur.
In 1D S=$\frac{1}{2}$ magnetic chain compounds, CuGeO$_3$ is a
prominent example \cite{Hase:1993}, a structural phase transition
into a statically dimerized structure is observed well above the
magnetic ordering temperature. The nature of this
spin-Peierls-transition is purely quantum mechanical. The pairing
of neighboring antiferromagnetic $S=\frac{1}{2}$ spins to a $S=0$
singlet leads to an energy gap thereby suppressing thermal spin
fluctuations and lowering the ground state energy.

\begin{figure}
\centering
\resizebox{1.00\columnwidth}{!}{\rotatebox{0}{\includegraphics{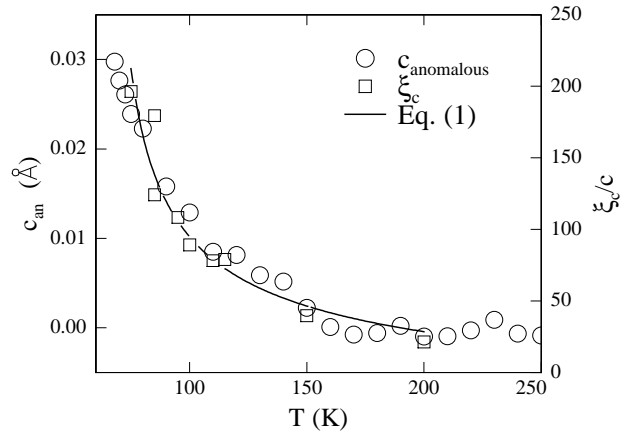}}}
\caption[]{Temperature dependence of the anomalous part of the
lattice constant $c$ ($\bullet$, see text for explanation) and the
1D correlation length $\xi_c$ ($\square$). The solid line is the
fit from Fig~\ref{kohaerpics} after
Eq.~(\ref{korrelationslaenge1}).} \label{critfluc}
\end{figure}

What is the reason for the unusual temperature behavior of the 1D
lattice parameter $c$? Fig.~\ref{critfluc} shows the anomalous
part of the temperature dependence of the lattice parameter,
i.e.\@ the difference of the lattice parameter $c$ from
Fig.~\ref{gitterkonstanten} b) to an extrapolated linear decrease
with temperature and the measured and calculated correlation
length $\xi_c$ from Fig.~\ref{kohaerpics} in the same plot.
When choosing the right scaling factor, the anomalous part of $c$,
$c_{an}$, and the correlation length $\xi_c$ fall on the same
curve and both are well described by the quasi 1D Ising model in
Eq.~(\ref{korrelationslaenge1}). That suggests a coupling of the
1D fluctuations in the paraelectric regime to local strains. Such
a coupling, however, is forbidden in the centrosymmetric space
group R$\bar{3}$; the coefficients of the (direct) piezoelectric
effect all vanish \cite{Nye:1979}. The peculiarity of the present
system is the unusual large amplitude of the paraelectric
fluctuations. Because of the reduced dimensionality of the
interactions, the ferroelectric fluctuations are strongly
enhanced. The anomaly of $c$ occurs below about 170~K, where the
correlation length already reaches about 40 lattice units,
$\xi_c(170\mbox{ K})/c\simeq 40$ (see Fig.~\ref{kohaerpics}).
Coupling via higher order terms in susceptibility and strain,
which become important due to the long ranged correlations, then
presumably leads to the observed electrostriction. This effect is
likely to occur in other low-dimensional systems, but has not been
reported before. We do not know, why the electrostriction leads to
a dilatation of the $\beta$-hydroquinone host lattice ($c_{an}>0$)
and a larger distance of the pseudospins in the 1D direction.
Intuitively, a contraction along $c$ should be energetically more
favorable.

\begin{figure}
\centering
\resizebox{1.00\columnwidth}{!}{\rotatebox{0}{\includegraphics{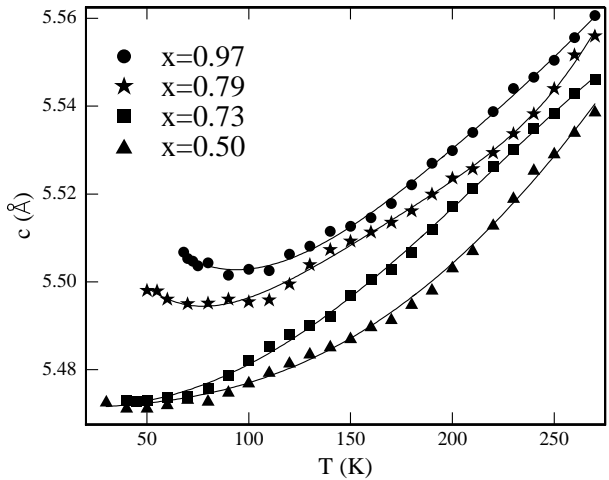}}}
\caption[]{Hexagonal lattice parameter $c$ for the concentrations
x=0.97, x=0.79, x=0.73 and x=0.50. The curves are shifted
vertically, for clarity. At 270~K, the lattice parameters for all
concentrations coincide.} \label{allec}
\end{figure}

We investigated methanol-clathrates with methanol-occupancies less
than 100~\% to prove that the anomaly of $c$ is a true property of
the EDD interaction. By lowering the fractional occupancy of the
cavities, the interaction between the pseudospins is diluted and
the transition into a dielectrically ordered state shifts to lower
temperatures. Samples with a larger percentage $x$ of filled
cavities ($x>x_c\simeq 0.76$) show conventional ordering via a
first order phase transition into an antiferroelectrically ordered
structure, whereas the others ($x<x_c$) freeze into dipolar
glasses \cite{Woll:2000,Woll:2001,Kityk:2002,Rheinstaedter:2003}.
The temperature dependence of the 1D lattice parameter $c$ for
several filling degrees down to 50~\% is shown in Figure
\ref{allec}.
Comparing the structural data with the static dielectric constant
for different degrees of occupation, see e.g.\@ Fig.~13 from
Ref.~[\onlinecite{Woll:2001}], there is an analogous behavior: The
anomaly is most pronounced for the completely filled
methanol-clathrate. When reducing the filling, the anomaly shifts
to lower temperatures and broadens.

\section*{Conclusion}\label{Conclusion}
The methanol clathrate can serve as a model system for dielectric
ordering and fluctuations and is conceptually close to magnetic
spin systems. Because of the lattice symmetry and the peculiar
property of the EDD interaction, the latter clathrate has quasi 1D
dielectric properties. From analysis of the superlattice
reflections in the dielectrically ordered phase, we can confirm
the theoretically predicted 4:2 structure. The latter ground state
can be calculated from Monte Carlo simulations and first
principles, considering the EDD interactions between the guest
molecules, only. We observe 1D and 3D critical scattering in the
paraelectric regime in neutron and X-ray diffraction experiments.
The temperature dependence of the correlation lengths $\xi_c$ and
$\xi_a$ is well described by the corresponding quasi 1D Ising
theory and agrees perfectly with results from dielectric
investigations, thereby confirming the microscopic picture of low
dimensional ordering and fluctuations of the methanol pseudospins
in the hydroquinone cagework. The EDD interaction on a deformable
lattice should be unstable against a dimerization of the dipoles
in the 1D chains. At the dielectric ordering temperature T$_c$ we
find superlattice reflections that indicate a doubling of the unit
cell in the 1D direction, i.e.\@ a static dimerization of the
dipole chains. We find no sign of diffuse scattering in
half-integer layers of $l$ due to fluctuations of the
dimerization. Below T=170 K an anomaly in the temperature
dependence of the 1D lattice parameter $c$ occurs. The large
amplitude of the 1D fluctuations in the paraelectric regime leads
to a coupling of the susceptibility to local strains and a
piezoelectric effect in the otherwise centrosymmetric space group
R$\bar{3}$. This effect is likely to be observed in low
dimensional interacting systems because fluctuations are strongly
enhanced and long ranged.

\acknowledgements We thank B.~Grenier and F.~Bourdarot for support
at the D23 diffractometer.

\bibliography{peierls}

\end{document}